\documentclass[12pt,a4paper]{article}
\usepackage[utf8]{inputenc}
\usepackage[T1]{fontenc}
\usepackage{amsmath,amssymb}
\usepackage{authblk}
\usepackage[round]{natbib}
\usepackage[noabbrev]{cleveref}
\usepackage{graphicx}
\usepackage[normalem]{ulem}
\usepackage{tikz}
\usepackage{caption}
\usepackage[hyphens]{url}
\title{Data Assimilation for a Geological Process Model Using the Ensemble Kalman Filter}
\author[]{Jacob Skauvold}
\author[]{Jo Eidsvik}
\affil[]{Department of Mathematical sciences, NTNU, Norway}

\begin{document}
\maketitle

\begin{abstract}
We consider the problem of conditioning a geological process-based computer simulation, which produces basin models by simulating transport and deposition of sediments, to data. Emphasising uncertainty quantification, we frame this as a Bayesian inverse problem, and propose to characterize the posterior probability distribution of the geological quantities of interest by using a variant of the ensemble Kalman filter, an estimation method which linearly and sequentially conditions realisations of the system state to data.

A test case involving synthetic data is used to assess the performance of the proposed estimation method, and to compare it with similar approaches. We further apply the method to a more realistic test case, involving real well data from the Colville foreland basin, North Slope, Alaska.
\end{abstract}

\section*{Introduction}
Process-based geological models are important for exploring connections between geological variables in a theoretical setting. The potential predictive value of the process-based approach has begun to receive recognition, but effective prediction requires that the model can be conditioned to observations. Conditioning methods for process-based models are typically impractical relative to data conditioning in other modelling settings, such as more traditional geostatistical models. Hence, examples of successful predictive application of process-based models are rare \citep{Pyrcz2014}.

This paper considers the problem of data assimilation for a geological process computer simulation, referred to as the Geological Process Model (GPM), where we specifically use the simulator developed by \citet{Tetzlaff2005}, which produces basin models by simulating transport and deposition of sediments, and erosion of existing geological layers. By data assimilation we mean bringing together information from well or seismic data and from the geological model, in a consistent manner, such that the result correctly characterises our knowledge about the system state--the geological details of the area under study--as well as other relevant parameters describing the depositional environment.

In this paper, data assimilation for the GPM is carried out using the ensemble Kalman filter (EnKF). In this filter, realisations of the model state, referred to as ensemble members, represent a sample from the probability distribution of the geological state variables. When observations, such as measurements of part of the actual geological system are to be assimilated, the simulation is halted and each ensemble member is modified to better match these observations. Then the simulation is resumed on the basis of the updated ensemble. The end result of completing the simulation, and assimilating all data, is a final updated ensemble which represents the posterior probability distribution of the geological quantities of interest given all available data. This is the desired solution to the data assimilation problem \citep{Evensen2009}.

There has been recent interest in uncertainty quantification and data conditioning for complex geological models. Promising approaches include the ones due to \citet{Charvin2009}, who use an iterative Monte Carlo sampling scheme to condition a 2D simulation of a shallow-marine sedimentation process to observations of thickness and grain size, \citet{Bertoncello2013}, who condition a surface-based model with iterative matching of sub-problems for a turbidite application, and \citet{Sacchi2015}, who use a mismatch criterion for well log and seismic data from simulations. By assimilating data gradually, the approach taken in the current paper exploits the way that the simulated sedimentation process forms layers in sequence. In cases where it is applicable, it has the potential to be considerably more efficient than other methods.

The next section provides a more detailed description of the GPM simulator, its inputs and outputs, and how the model state is represented. The subsequent \emph{Methodology} section gives an overview of the EnKF, and how it is implemented to work with the GPM (details are in the Appendix). In the \emph{Numerical experiments} section, the EnKF/GPM combination is tested on two different data sets: One synthetic case created using the GPM, and one real case with well data from North Slope, Alaska. We close with a discussion section, reviewing the strengths and weaknesses of the proposed data assimilation scheme in light of the results from the two test cases, and pointing out possible directions for further development.

\section*{Geological Process Model}

During the last three decades, the field of stratigraphic and sedimentological process modelling has seen much development, with simulation efforts including SEDSIM \citep{TetzlaffHarbaugh1989}, SEDFLUX \citep{SyvitskiHutton2001,HuttonSyvitski2008}, BARSIM \citep{Storms2003} and FLUMY \citep{Lopez2009}. See \cite{Paola2000} or \citet{TetzlaffPriddy2001} for details.

Process-based geological models differ from other geological and geostatistical models in that they seek to capture not only the nature of geology existing today, but also the processes which formed it. Process-based models, sometimes called process-response models, are powerful tools for establishing relationships between processes and results, especially when the processes in question cannot be simulated by a physical experiment in a laboratory. On the other hand, we require validation using field measurements or experimental observations in order to have confidence in process-based models, as well as any inferences drawn on the basis of their output.

One rather indirect way of using process-based models for prediction is to use the process simulation output as training data for some geostatistical prediction method, like multiple point statistics \citep{Edwards2016,Strebelle2002}. In doing this, one assumes that the process realisations used as training data are representative of the spatial structure of the geological features of interest so that, for instance, the variability in shape and size of features produced by simulation matches the variability observed in nature. One further assumes that the geostatistical method is able to extract the relevant structural information from the training data, and that this information generalises well enough to the geology of the prediction target. In other words, this ``digital analogue'' way of using process-based models to inform prediction requires essentially the same fundamental assumptions as do traditional geostatistical methods \citep{Pyrcz2014}. By constraining the simulation one avoids these assumptions, introducing in their stead the assumption that the process-based model is valid.

\begin{figure}
\begin{center}
\includegraphics[width=0.8\textwidth]{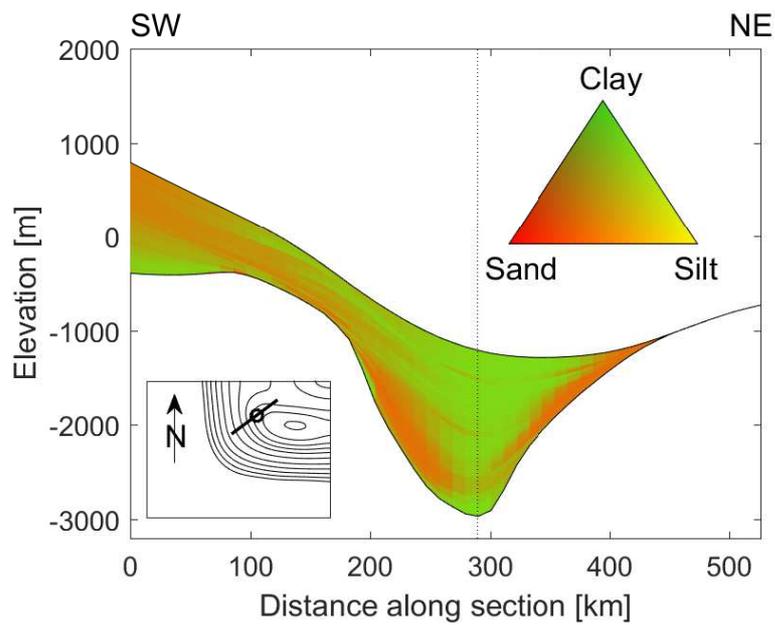}
\caption{Cross section of a simulated layer package with colours indicating proportions of sediment types (sand, silt, clay) in each position. The elevations and sediment proportions shown here are the ensemble mean of the final analysis ensemble in the North Slope, Alaska test case. The dotted vertical line near the centre indicates the location of the Tunalik 1 well. Inset map shows locations of section and well in modelled basin.}
\label{fig:2Dsection}
\end{center}
\end{figure}

The GPM considered in this paper produces basin models by simulating transport and deposition of sediments, and erosion of existing geological layers \citep{Tetzlaff2005, Christ2016}. The same software is also capable of simulating other processes, such as carbonate growth, though that is not discussed in this paper. The basin is filled by sediments entering at a defined source location. In this case there is no sink in the model, and a gradual basin-filling process takes place, where the layer composition is defined by the sediment supply at the source and the sea level. The composition of particles in the sediment supply is kept fixed in our case, but since it can be modified in the software, it could be included in the statistical model as a set of uncertain parameters to be estimated from data. An overview of the simulator's input and output variables is given in Table \ref{tab:inputoutput}. The graph just below the table illustrates the relation between input and output. We next discuss each element in more detail. Figure \ref{fig:2Dsection} shows an example of model output from the GPM conditioned to data (the context will be clarified in the examples).

\begin{table}
\begin{center}
\caption{Input and output variables of the GPM simulator. The dimensions $n_x$ and $n_y$ define the size of the horizontal grid, and $n_t$ is the number of discrete time steps used, starting at geological time $t_0$ and moving forward until time $t_{n_t}$. Some input variables have no symbol as they are not modelled explicitly in this paper.}
\label{tab:inputoutput}
\begin{tabular}{llll}
\hline
	& Variable	& Symbol & Dimensions, Type \\
\hline
\textbf{Input} & Initial bathymetry & $\mathbf{z}_0$ & $n_x \times n_y$ matrix \\
	& Sea level curve & $\boldsymbol\theta_\mathrm{SL}$ & $n_t \times 1$ vector \\
	& Sediment supply rate & $\boldsymbol\theta_\mathrm{SS}$ & $n_t \times 1$ vector \\
	& Sediment source locations & - & $n_x \times n_y$ matrix \\
	& Tectonic uplift/subsidence rate & - & $n_x \times n_y$ matrix \\
	& Initial surface sediment proportions & - & $4\times 1$ vector \\
\hline
\textbf{Output} & Surface elevation & $\mathbf{z}_k$ & $n_x \times n_y$ matrix \\
	& ($k$th layer, $k = 1,\ldots, n_t$) & & \\
	& Sediment proportion & $\mathbf{p}_{k,l}$ & $n_x \times n_y$ matrix \\
	& ($k$th layer, $l$th sediment type) & & \\
\hline
\end{tabular}
\end{center}
\begin{center}
\begin{tikzpicture}
\node [draw, circle, anchor=east] (x0) at (-1,0) {$\mathbf{x}_0$};
\node [draw, circle, anchor=east] (theta) at (-1,-1) {$\boldsymbol\theta$};
\node [draw] (F) at (0,0) {$F$};
\node [draw, circle, anchor=west] (x1) at (1,0) {$\mathbf{x}_1$};

\node [anchor=east] (z0p0) at (-2,0) {$\mathbf{x}_0 = (\mathbf{z}_0, \mathbf{p}_0)$};
\node [anchor=west] (z1p1) at (2,0) {$\mathbf{x}_1 = (\mathbf{z}_1, \mathbf{p}_1)$};
\node [anchor=east] (thetaSLSS) at (-2,-1) {$\boldsymbol\theta = (\boldsymbol\theta_\text{SL}, \boldsymbol\theta_\text{SS})$};

\draw[->,>=latex] (theta) -- (F);
\draw[->,>=latex] (x0) -- (F);
\draw[->,>=latex] (F) -- (x1);
\end{tikzpicture}
\end{center}
\end{table}

The forward model $F$ represents the GPM simulation software. We treat $F$ as a black box which accepts as input the system state $\mathbf{x}_k$ at time $t_k$, and the vector $\boldsymbol\theta$ of environmental parameters, and returns the system state $\mathbf{x}_{k'}$ at a later time $t_{k'} > t_k$. We are free to choose the time interval $t_{k'} - t_k$, but can only go forward in time. The forward model $F$ is deterministic. Given the same input, it will always produce the same output. We thus make the important assumption that $F$ is a correct representation of reality in the sense that there is no stochastic model error associated with it. The state uncertainty will be represented by an ensemble; multiple input realisations at time $t_k$ are propagated through $F$ to give multiple output representations at time $t_{k'}$.

The parameter vector $\boldsymbol\theta$ consists of the sea level curve $\boldsymbol\theta_\text{SL}$, which describes the evolution of the sea level over simulated geological time, and the sediment supply curve $\boldsymbol\theta_\text{SS}$ which specifies, as a function of time, the rate at which sediment enters the model area from outside (sediment source locations must also be specified, but the details of this will not be considered here). The sea level and sediment supply curves are represented as piecewise linear functions over time, with the vectors $\boldsymbol\theta_\text{SL}$ and $\boldsymbol\theta_\text{SS}$ containing function values at a shared set $\{t_i: i = 0, \ldots, n_t \}$ of discrete time points. Other quantities could have been included as parameters, such as the intensity of erosion as a function of water depth, or the rate of tectonic uplift and subsidence as a function of time and horizontal location. In the interest of a limited scope, however, we have chosen to focus on the sea level and sediment supply curves in this study. Other parameters which are required input for the simulator, are treated as known quantities, and kept fixed throughout.

The state vector $\mathbf{x}$ represents the physical configuration of the modelled system at a given moment in time. Together with the parameter vector $\boldsymbol\theta$, it contains all the information necessary to compute the system state at a later time. The rationale for treating $\mathbf{x}$ and $\boldsymbol\theta$ as separate entities is the asymmetry of the causal relationship between them; namely that the parameters influence how the state evolves over time, but the state does not influence the parameters.

There are two components of the state vector $\mathbf{x}$: The elevation component $\mathbf{z}$ and the sediment proportion component $\mathbf{p}$, which specifies how much sediment belongs to each of the categories coarse sand, fine sand, silt, and clay. Both are defined over a two-dimensional grid of discrete locations. (Additional details about the state vector are given in the Appendix.)

The elevation component $\mathbf{z}$ is a set of surfaces corresponding to the boundaries between layers of sediment deposited during successive time steps. The initial state $\mathbf{x}_0$ has only one elevation surface, $\mathbf{z}_0$, referred to as the initial bathymetry. After running the simulator $F$ from time $t_0$ to time $t_1$, there will be two elevation surfaces, $\mathbf{z}_0$ and $\mathbf{z}_1$, corresponding to the bottom and top of the layer formed during the time interval $(t_0, t_1)$. Due to erosion and tectonics, the new bottom surface $\mathbf{z}_0$ at time $t_1$ will generally be different from the original bottom surface $\mathbf{z}_0$ at time $t_0$.

The sediment proportion component $\mathbf{p}$ is a field of proportions characterising the type of sediment present in each location. To each three-dimensional grid cell $c$---defined horizontally by the two-dimensional model grid, and bounded vertically by two successive elevation surfaces in $\mathbf{z}$---is associated a vector $\mathbf{p}(c) = (p_1(c), p_2(c), p_3(c), p_4(c))$ specifying the proportion of each of four discrete sediment types (coarse and, fine sand, silt, and clay) present in the cell. In the following, the cell index $c$ will be suppressed from the notation when the meaning is clear from the context.

\section*{Methodology}

\subsection*{Uncertainty quantification}

To meaningfully characterise quantitative geological variables of interest, it is necessary to assess the uncertainty associated with predictions and parameter estimates. For non-linear systems it is natural to quantify uncertainty with a sample or an ensemble of Monte Carlo realisations. As mentioned above, the GPM forward model $F$ is assumed to be deterministic, and realisations of the geological system state are obtained by propagating sampled initial conditions forward in time under different versions of the sea-level and sediment supply parameters, which are also drawn from their prior distributions. The resulting output of GPM is an ensemble of state variables at relevant geological times. As this ensemble is purely model-driven and has not been conditioned to data, we refer to it as a prior ensemble.

In its Bayesian flavour, the approach described in \citet{Charvin2009} is quite similar to that of the current paper. But unlike their Markov chain Monte Carlo sampling procedure for assessing the posterior probability density function (pdf), the EnKF approach described here performs sequential updating of the state variables.

\subsection*{EnKF conditioning approach}

\begin{table}
\begin{center}
\caption{Generic variables and model components in the hidden Markov model  (HMM) view of dynamical processes, and corresponding entities in the GPM setting. The graph illustrating the HMM dependence structure of the state and observation variables at discrete time points $t_0, t_1, \ldots, t_{n_t}$ as well as the parameter vector $\boldsymbol\theta$. Arrows between nodes indicate statistical dependence.}
\label{tab:graph}
\begin{tabular}{ll}
\hline
Generic         & GPM-specific \\
\hline
System state $\mathbf{x}$ 	& Layer elevation $\mathbf{z}$ \\
				& Layer sediment composition $\mathbf{p}$ \vspace{5pt}\\
Initial state $\mathbf{x_0}$ 	& Initial bathymetry $\mathbf{z}_0$ \\
				& Base layer sediment composition $\mathbf{p}_0$ \vspace{5pt}\\
Parameters $\boldsymbol\theta$ & Sea level $\boldsymbol\theta_\text{SL}$ \\
				& Sediment supply $\boldsymbol\theta_\text{SS}$ \vspace{5pt}\\
Dynamic model $F$	& Geological process simulator \vspace{5pt}\\
Observations $\mathbf{y}$ 	& Well logs \vspace{5pt}\\
Observation model $\mathbf{h}$ & Synthetic well logs \\
\hline
\end{tabular}
\end{center}
\begin{center}
\begin{tikzpicture}
\def \dx {1.4cm}
\def \dy {1.4cm}

\node [draw,circle] (x0) at (0,0) {$\mathbf{x}_0$};
\node [draw,circle] (x1) at (\dx,0) {$\mathbf{x}_1$};
\node [draw,circle] (x2) at (2*\dx,0) {$\mathbf{x}_2$};
\node [] (cdots) at (3*\dx,0) {$\cdots$};
\node [draw,circle] (xT) at (4*\dx,0) {$\mathbf{x}_{n_t}$};

\node [draw,circle] (y1) at (\dx,\dy) {$\mathbf{y}_1$};
\node [draw,circle] (y2) at (2*\dx,\dy) {$\mathbf{y}_2$};
\node [draw,circle] (yT) at (4*\dx,\dy) {$\mathbf{y}_{n_t}$};

\node [draw,circle] (theta) at (2*\dx,-\dy) {$\boldsymbol\theta$};

\draw[->,>=latex] (x0) -- (x1);
\draw[->,>=latex] (x1) -- (x2);
\draw[->,>=latex] (x2) -- (cdots);
\draw[->,>=latex] (cdots) -- (xT);

\draw[->,>=latex] (x1) -- (y1);
\draw[->,>=latex] (x2) -- (y2);
\draw[->,>=latex] (xT) -- (yT);

\draw[->,>=latex] (theta) -- (x1);
\draw[->,>=latex] (theta) -- (x2);
\draw[->,>=latex] (theta) -- (xT);
\end{tikzpicture}
\end{center}
\end{table}

The graph in Table \ref{tab:graph} illustrates the geological variables as a latent process. The variables are coupled in time according to the GPM forward model $F$, which is assumed to be Markovian in the sense that only the current state is relevant to the future evolution of the system, not the history leading to the current state. The top nodes in the graph illustrate data on which the process simulation is conditioned. Here, we assume that the data are well log observations of elevations and sediment proportions at discrete intervals in geological time, although other sources of information could also be used, such as attributes derived from seismic data. We assume that the observations are made in one well, at grid coordinates $(i^{\text{obs}},j^{\text{obs}})$, for layers deposited at $n_t$ different time points; $t_1 < \ldots < t_{n_t}$. The synthetic data used in the simulation study have the form $\mathbf{y}_k=(z_k(i^{\text{obs}},j^{\text{obs}}),\mathbf{p}_k(i^{\text{obs}},j^{\text{obs}}))$, while the data in the Alaska North Slope case consist of thickness and gamma ray observations linked to sediment proportions.

The well log data are modelled by a likelihood function. This means that a conditional pdf for the data is specified given the geological variables. Conditional on the geological elevation and sediment proportion at a time $t_k$, the measurement has an expectation defined by a functional relationship $\mathbf{h}(\mathbf{x}_k)$ and an additive, zero-mean Gaussian noise $\mathbf{\epsilon}_y$ with covariance $\text{Cov}(\mathbf{\epsilon}_y,\mathbf{\epsilon}_y)$. The observation operator $\mathbf{h}$ could simply select values of the state variables (here elevation and sediment proportions) at the observation site, i.e. the location of the well in the model grid. In realistic settings however, $\mathbf{h}$ will typically have a more complex form, such as a local spatial average or a nonlinear function of one or more state variables. Such operators may involve parameters which require tuning to provide an adequate likelihood model for a specific application. The observation operator for gamma ray measurements used in the North Slope, Alaska case is an example of this.

In real applications the likelihood model will also require some form of matching between simulated depth at the time of deposition and measured depth at the time of observation. The data $\mathbf{y}_1, \ldots, \mathbf{y}_{n_t}$ shown in the graph in Table \ref{tab:graph} are assumed to be informative of the system state at the time of deposition. In the synthetic simulation study, this matching problem is avoided altogether by recording synthetic observations during, rather than after, the simulation. For the North Slope, Alaska case, a time-to-thickness relationship is established ahead of time as a part of model calibration. This involves sampling initial states and parameters from the prior distribution, and running the simulation based on these without assimilating data. The resulting unconditional model runs are used to construct a time-to-depth curve.

The goal of data assimilation is to characterise the posterior pdf of the system state, given all data by the current time step: $(\mathbf{y}_1,\ldots,\mathbf{y}_k)$. In the EnKF, the solution is constructed sequentially; forecasting one step ahead using the GPM model $F$, and then conditioning on one more part $\mathbf{y}_k$ of the data, at every time step $k$. It is convenient for conditioning purposes to build an augmented state vector that includes geological layer variables for all previous geological times. The sea level and sediment supply parameters are also part of this augmented state vector. These parameters are distinct from the geological layer variables only in the sense that they are not changed by the GPM forward model. Hence, the geological state variables change in both the forecast and update steps, while the parameters change only in the update step. 

The details of the EnKF implementation are provided in the Appendix, but the important elements are summarised here. To apply the EnKF to the GPM data assimilation problem, $n_e$ samples from the initial state $\mathbf{x}_0$ and parameters $\boldsymbol\theta$ are generated from the prior pdf. Then the GPM runs from time $t_0$ until $t_1$ for all $n_e$ ensemble members, giving an $n_e$-member forecast ensemble at time $t_1$. Using a generic notation where $\mathbf{v}_1$ denotes the state vector after one time step, the forecast ensemble at that time is
\[
\mathbf{v}_1^{1,\text{f}},\mathbf{v}_1^{2,\text{f}},\ldots,\mathbf{v}_1^{n_e,\text{f}}.
\]
Next, for each ensemble member $b$, pseudo-data are created by evaluating the expectation in the likelihood $\mathbf{h}_1(\mathbf{v}_1^{b,\text{f}})$ and adding a random Gaussian perturbation $\boldsymbol\epsilon_{y,1}^b$ with the likelihood covariance. Thus, the pseudo-data are
\begin{equation}\label{pseudo_y}
\mathbf{y}_1^b = \mathbf{h}_1(\mathbf{v}_1^{b,\text{f}}) + \boldsymbol\epsilon_{y,1}^b.
\end{equation} 
Finally, the Kalman filter update is applied to each ensemble member $b = 1, \ldots, n_e$,
\begin{eqnarray}\label{kf_eq}
\mathbf{v}_1^{b,\text{a}} &=& \mathbf{v}_1^{b,\text{f}} + \hat{\mathbf{K}}_1 \left(\mathbf{y}_1 - \mathbf{y}_1^b\right), \\
\hat{\mathbf{K}}_1 &=& \widehat{\mbox{Cov}}[\mathbf{v}_1^\text{f},\mathbf{h}_1(\mathbf{v}_1^\text{f})] \left( \widehat{\mbox{Cov}}[\mathbf{h}_1(\mathbf{v}_1^\text{f}),\mathbf{h}_1(\mathbf{v}_1^\text{f})] + \mbox{Cov}(\boldsymbol\epsilon_{y},\boldsymbol\epsilon_{y}) \right)^{-1}. \nonumber
\end{eqnarray}
The covariances are estimated empirically from the forecast ensemble (see Appendix). Once all ensemble members have received their respective updates, an analysis ensemble
\[
\mathbf{v}_1^{1,\text{a}},\mathbf{v}_1^{2,\text{a}},\ldots,\mathbf{v}_1^{n_e,\text{a}}.
\]
of size $n_e$ is available after time step 1.

This forecast-update cycle is then repeated, using the newly formed analysis ensemble instead of the prior ensemble used initially, producing first a $t_2$-forecast ensemble, then a $t_2$-analysis ensemble, and so on. With each update, data from one observation vector is integrated into the ensemble. In probability density terms, the conditional pdf of $\mathbf{v}_2$ given $\mathbf{y}_1$ and $\mathbf{y}_2$ is $p(\mathbf{v}_2|\mathbf{y}_1,\mathbf{y}_2) \propto p(\mathbf{y}_2|\mathbf{v}_2)p(\mathbf{v}_2|\mathbf{y}_1)$, where the first term on the right hand side is the likelihood model of $\mathbf{y}_2$, which is conditionally independent of the other variables given $\mathbf{v}_2$ (see dependence structure in Table \ref{tab:graph}). The second term is the forecast pdf which is represented by taking each ensemble member from the previous time step forward one step using the GPM. When all data have been assimilated into the analysis ensemble at the last time point $t_{n_t}$, the ensemble is representative of the posterior pdf of all geological variables, given all the data.

In the simulation study below, this EnKF approach is compared with two alternative methods. 
The first is often called the Ensemble Kalman Smoother (EnS), see e.g. \cite{Evensen2009}. It runs the ensemble members forward through all time steps before updating. 
The benefit of this approach in the geological process setting is that data are compared at the same geological time, which makes the likelihood model easier to specify.
The downside is that it is very difficult to match present-day geology directly. In contrast, a filtering approach which integrates data sequentially is guided towards more realistic solutions as it steps forward through geological time.

The second alternative approach is EnS with multiple data assimilation (MDA), as described by \citet{emerick2013}. The MDA approach relies on the following relation between pdfs: 
$p(\mathbf{y}_k|\mathbf{v}_k) = \prod_{r=1}^R p(\mathbf{y}_k|\mathbf{v}_k)^{1/R}$.
Just like EnS, the ensemble members are now run all the way through the geological time interval, and updating is done at the end. But the MDA approach runs forward $R$ times, with each update using an inflated likelihood covariance, $R$ times larger than the actual observation covariance. A larger covariance means that the linear updates are smaller than the ones in the EnS. It can be difficult to tune $R$ in practice, and if an application calls for a large number of iterations, the computational cost will be high.

\section*{Numerical experiments}
\subsection*{Synthetic data}
\begin{figure}
\begin{center}
\includegraphics[width=\textwidth]{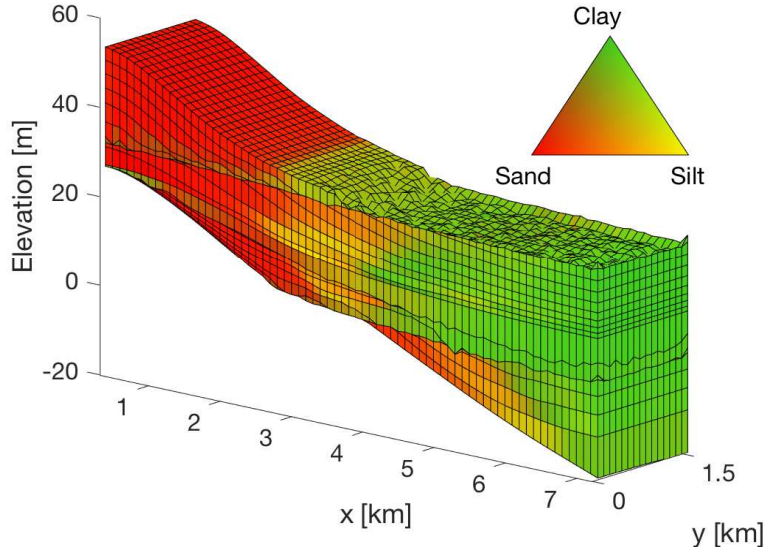}
\caption{An example of a reference realisation created with GPM.}
\label{fig:referenceRealization}
\end{center}
\end{figure}

To demonstrate how the data conditioning works in practice, we apply it to an artificial test case. Our case is inspired by, but distinct from, the case considered by \citet{Charvin2009}. A reference realisation has been created by simulating sediment diffusion over 20 000 years. We use a grid consisting of $n_x = 72$ by $n_y = 16$ cells. Each cell has a horizontal size of 100 by 100 meters, so that the modelled region is a rectangle, 7.2 kilometers long in the cross-shore direction and 1.6 kilometers wide in the along-shore direction. The initial surface is roughly planar, with a downward slope of approximately $0.4^\circ$ in the positive $x$-direction, but it also has smoothly varying deviations from this trend, drawn from a Gaussian random field with a correlation range of 10 grid cells, or 1 kilometer. Sediments enter the modelled area along the landward edge of the grid, at the top of the slope. New sediments appear here at a rate controlled by the sediment supply parameter $\boldsymbol\theta_\mathrm{SS}$. Over time they diffuse downhill, and are deposited at various distances down slope, depending on grain size. Figure \ref{fig:referenceRealization} shows an example of a realisation, at the final time of the simulation.

\begin{figure}
\begin{center}
\includegraphics[width=\textwidth]{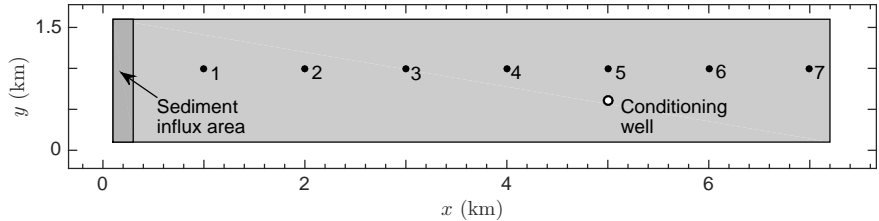}
\caption{Locations of the conditioning well and the 7 blind wells in the modelled area. The source area at the top of the slope, where new sediment is introduced, is also indicated.}
\label{fig:bwlocations}
\end{center}
\end{figure}

We assume that data are layer elevations and sediment proportions from a vertical well. The location of this conditioning well is shown in Fig. \ref{fig:bwlocations}. The goal is then to recover the layer package of the reference realisation by conditioning new GPM simulations to the data from this well. The initial bathymetry, sediment supply rate and sea level curve are all considered unknown, and must be estimated. The setting is an ideal case for the filtering method in the sense that the ensemble members are generated by the same process as the reference. Thus, model misspecification is essentially eliminated as a source of error. Any observed mismatch between the reference case and the filtering prediction will be due to limitations of the methodology, and not the simulation model itself. This is not the case when working with real data.

We refer to the experiment of generating a GPM reference realisation, assigning well log data, and predicting the model state from this data, as one trial. To assess the performance of the filtering method, 100 independent trials were performed. Results of each trial, including both the reference realisation and the prediction, are stored for seven different ``blind wells'' placed at regular intervals down the length of the modelled area. The positions of the blind wells in relation to the conditioning well are indicated in Fig. \ref{fig:bwlocations}.

From the 100 trial results, we compute the following statistics to gauge the quality of the predictions obtained from the ensemble representation in the EnKF: 
\begin{itemize}
    \item Mean square error (MSE) which measures the average square difference between reference realisation and prediction. Smaller values of MSE mean better prediction.
    \item Continuously ranked probability score (CRPS) which measures the accuracy of the predictive distribution represented by the ensemble, relative to the reference blind well data \citep{Gneiting2007}. Smaller CRPS is preferred since it means more precise predictions.
    \item Empirical coverage probability (Cov.Pr.) of 80\% confidence intervals, which is the empirically observed proportion of trials producing confidence intervals which actually cover the corresponding value of the reference realisation. A probability near 80\% means the ensemble members correctly quantify the uncertainty associated with the prediction. With an ensemble size of 100 it is convenient to form an 80 percent confidence interval by trimming 10 ensemble members from each tail of the distribution.
\end{itemize} 

The results are given in Table \ref{tab:EnKFtrialresults}, where we averaged over all 20 layers.
\begin{table}
\begin{center}
\caption{EnKF trial results for $z$ and $\mathbf{p}$.}
\label{tab:EnKFtrialresults}
\begin{tabular}{llrrrrrrr}
\hline
    &           &       &       &   \multicolumn{3}{c}{Well number} & & \\
    &           &   1   &   2   &   3   &   4   &   5   &   6   &   7   \\
\hline
$z$ & MSE       & 18.96	& 13.55	& 14.37	& 9.62	& 3.62	& 6.80	& 15.36	\\ 
    & CRPS      & 1.48	& 1.13	& 1.01	& 0.80	& 0.53	& 0.74	& 0.99	\\ 
    & Cov.Pr.(80)  & 0.67	& 0.70	& 0.70	& 0.70	& 0.73	& 0.72	& 0.72	\\ 
\hline
$\mathbf{p}$ & MSE       & 49.98	& 51.74	& 37.58	& 30.42	& 12.45	& 23.96	& 16.59	\\ 
    & CRPS      & 1.29	& 1.34	& 1.08	& 0.98	& 0.52	& 0.85	& 0.79	\\ 
    & Cov.Pr.(80)  & 0.77	& 0.73	& 0.74	& 0.76	& 0.84	& 0.76	& 0.73	\\ 
\hline
\end{tabular}
\end{center}
\end{table}
The smallest values of MSE and CRPS are the ones for blind well number 5, which is closest to the conditioning well. This holds both for the layer depths and for the proportions. It is more difficult to predict far from the conditioning well. The coverage probabilities tend to be a little below 80\%, but no spatial trends are apparent. Nor is there any dramatic underestimation of the uncertainty. 

Next, we compare the EnKF approach with EnS and MDA. Summary statistics over 100 trials are given in the Table \ref{tab:trialresults}.
In this case, both the EnS and MDA are clearly underestimating the uncertainty, which lessens the quality of the predictions significantly. The MDA used $R = 4$ iterations, which is not necessarily optimal, but it is not obvious how the number of iterations $R$ should be tuned.

The statistics reported for $\mathbf{p}$ in Table \ref{tab:EnKFtrialresults} and Table \ref{tab:trialresults} are not computed directly from the proportion vector itself, but a different variable $\mathbf{s}$ related to $\mathbf{p}$ via a logistic transformation. The reason for using a transformation is that while the elements of $\mathbf{p}$ must be valid probabilities, the elements of $\mathbf{s}$ can take any real value. See the Appendix for details.
\begin{table}
\begin{center}
\caption{Trial results for EnKF, EnS and MDA compared in terms of MSE, CRPS and coverage probability.}
\label{tab:trialresults}
\begin{tabular}{llrrr}
\hline
    &       & EnKF 	& EnS & MDA \\
\hline
$z$	& MSE 	& 11.76 & 978.50 & 996.69 \\
	& CRPS 	& 0.95 & 22.98 & 24.93  \\
	& Cov. Pr. 	& 0.70 & 0.07 & 0.01  \\
\hline
$\mathbf{p}$ 	& MSE & 31.82 & 177.08 & 134.04 \\
	& CRPS & 0.98 & 3.88 & 2.76 \\
	& Cov. Pr. & 0.76 & 0.31 & 0.48 \\ 
\hline
$\theta_\text{SL}$ 	& MSE & 125.78 & 528.62 & 335.72 \\
		& CRPS & 4.56 & 12.43 & 12.00 \\
		& Cov. Pr. & 0.58 & 0.37 & 0.23 \\ 
\hline
$\theta_\text{SS}$ 	& MSE & 13.72 & 181.19 & 153.15 \\
		& CRPS & 1.42 & 8.57 & 9.66 \\
		& Cov. Pr. & 0.66 & 0.23 & 0.08 \\
\hline
\end{tabular}
\end{center}
\end{table}

\begin{figure}
\begin{center}
\includegraphics[width=\textwidth]{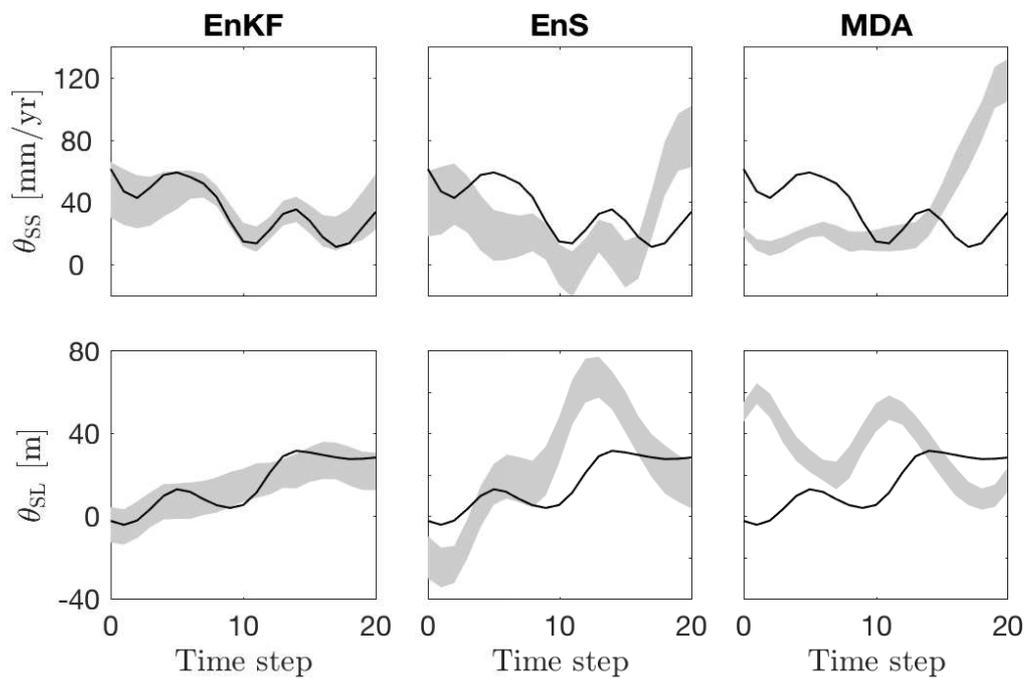}
\caption{Example trial results for the sediment supply (top) and sea level (bottom) parameters. The shaded areas indicate empirical 80\% confidence intervals constructed from ensemble members obtained from the EnKF (left), the EnS (middle) and MDA (right). The true parameter values for the trial in question are shown as solid lines.}
\label{fig:slsstrial_EnKF}
\end{center}
\end{figure}

Figure \ref{fig:slsstrial_EnKF} shows the posterior distributions for sea level and sediment supply as a function of geological time for a single trial. The EnKF results (left) are clearly better at covering the reference values. EnS (middle) and MDA (right) tend to give biased results. The sea level prediction obtained by EnKF only covers the larger geological time trends. Based on only one conditioning well, it appears difficult to capture the smaller fluctuations giving coarsening and fining upwards trends in Fig. \ref{fig:referenceRealization}.

\subsection*{Real data case: North Slope, Alaska}

\begin{figure}
\begin{center}
\includegraphics[width=0.8\textwidth]{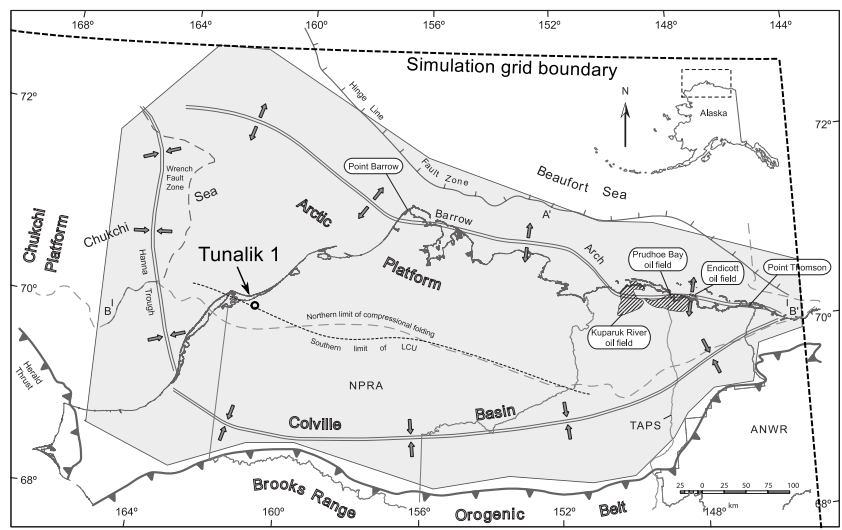}
\caption{Location of Tunalik 1 well relative to the Colville foreland basin study area (shaded region), and to the present-day coastline of northern Alaska. Parts of the northern and eastern boundaries of the simulation grid are shown as dashed lines. The southern and western boundaries are located outside the area covered by the figure. The map shown here is an adaptation of Fig. 1 in \cite{Schenk2012}}
\label{fig:polygon}
\end{center}
\end{figure}

The northern part of Alaska is an important oil and gas region, with much available data in the form of well logs and seismic surveys. In this section, we use GPM to model the Colville foreland basin. The area is indicated in Fig. \ref{fig:polygon}. This is the area studied by \citet{Schenk2012}. Starting with an initial bathymetry corresponding to the top surface at 120 Ma, we use GPM to simulate deposition in the basin until 115 Ma. The simulation is conditioned on gamma ray well log data from the Tunalik 1 well, located at $70.20^\circ \ \text{N}, 161.07^\circ \ \text{W}$, indicated by a circle on the map in Fig. \ref{fig:polygon}. The gamma ray (GR) data are measurements of natural gamma radiation taken at regular depth intervals along the trajectory of the borehole. It is measured in API (American Petroleum Institute) units, defined as a scaling of the observed radioactivity count rate by that recorded with the same logging tool in a reference depth zone~\citep{killeen1982gamma,keys1996practical}.

The lateral model grid covers a rectangular geographical area measuring approximately 1600 km in the east-west direction and 1300 km in the north-south direction, which is discretised into $110 \times 87$ grid cells, yielding a lateral resolution of ${\sim}15 \ \text{km}$ in either direction. The Colville foreland basin is located in the northeast corner of the rectangle (see Fig. \ref{fig:polygon}). The southern and western parts of the rectangle are included in order to properly model sediment entering the actual basin region.

\begin{figure}
\begin{center}
\includegraphics[width=\textwidth]{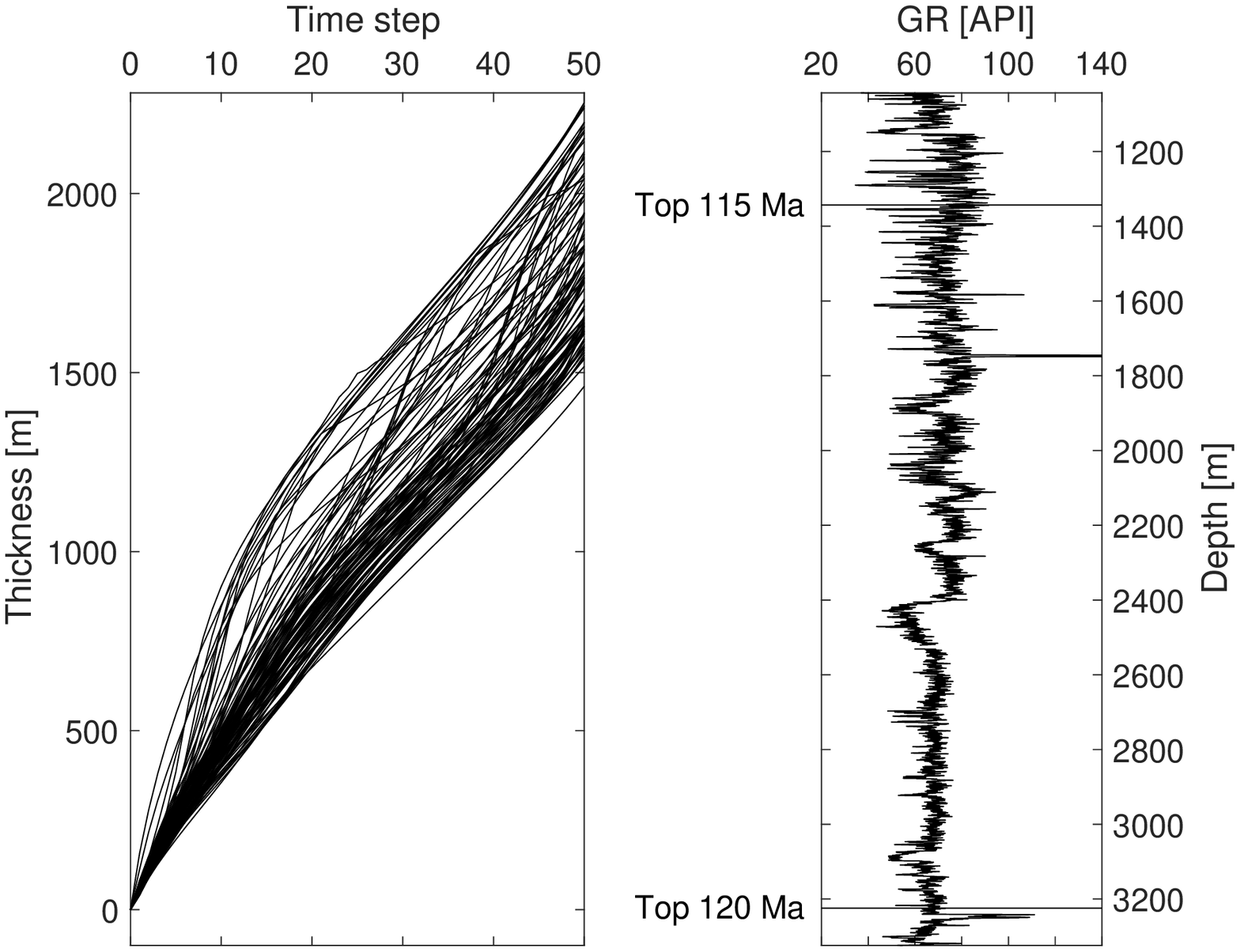}
\caption{Left: Prior realisations of cumulative thickness of deposited sediment at well location over simulated geological time. Right: The part of the Tunalik 1 gamma ray well log believed to be informative of the sedimentation happening between 120 Ma and 115 Ma.}
\label{fig:Tunalik1gammaRay}
\end{center}
\end{figure}

The Tunalik 1 well is approximately 6 km deep and, given the coarse horizontal spatial resolution of the model grid, can safely be assumed to be vertical. Based on the existing conceptual model of the study area by \cite{Christ2016}, the part of the well log relevant to the modelled time interval is believed to be the ${\sim}1900$ m depth interval between 1300 m and 3200 m of depth relative to the present-day surface. The right panel of Fig. \ref{fig:Tunalik1gammaRay} shows this part of the Tunalik 1 gamma ray log. The Tunalik 1 well data is available in LAS-format online \citep{USGS1981}.

The time interval between 120 Ma and 115 Ma is discretised into 50 time steps of $100 \ 000$ years each. Consequently, the completed model output will consist of 50 distinct layers. We choose to update the model once every fifth time step. This means that the forecast ensemble after 5 time steps consists of only the first 5 layers, and is updated based on the deepest part of the gamma ray log shown in the right panel of Fig. \ref{fig:Tunalik1gammaRay}. Similarly, after 10 time steps, each member of the forecast ensemble contains 10 layers, and they are all updated based on the next segment of the well log, and so on, at times $15, 20, \ldots, 50$. 

Time steps of $100 \ 000$ years were chosen as a compromise between model resolution, both temporal and spatial, on the one hand, and computational efficiency on the other hand. Using shorter time steps would produce a larger number of thinner layers, which would allow us to resolve smaller details. At the same time, limiting the number of layers by using longer time steps reduces the amount of information that has to be passed to and from the simulator during data assimilation, making the procedure more efficient. Based on our experience with the simulator, we believe that time steps of $100 \ 000$ years give us the resolution necessary to capture relevant changes in grain size over time, such as the progradational Brookian sequences we are trying to model in this case~\citep{Christ2016}.

The reason for assimilating data only once every five time steps, and observing blocks of five layers at a time, is that early experiments showed that updating on every time step tended to cause overfitting. That is, observations would be matched closely by the estimated system state, but the latter would have changed so much to accommodate the observations as to be unrealistic. At the other extreme, updating every tenth time step, and observing larger blocks of ten layers tended to produce very smooth estimates of the system state.

In order to carry out updates, it is necessary to identify which segment of the well log corresponds to the most recent five-layer block in the state forecast. This matching relies on a mapping between simulated time and thickness of the part of the well log relevant to the layers which have been simulated after that amount of time. In other words, a curve specifying cumulative present day thickness of the deposited layer package as a function of time. Rather than resorting to traditional back-stripping and decompaction methods, we obtain an estimate of this time-to-thickness map using results of unconditional simulations (i.e. model runs without data assimilation) carried out in advance, with initial state and environmental parameters drawn from the prior pdf. The left panel of Fig. \ref{fig:Tunalik1gammaRay} shows 100 realisations of the time-to-thickness relationship at the location of the Tunalik 1 well. The curves shown in the figure give the thickness at 115 Ma. Dividing each curve by its final thickness and multiplying by the thickness of the relevant well log depth interval gives standardised thickness curves. The specific map used for conditioning is a single sequence of depths $\{\Delta z_0, \Delta z_1, \Delta z_2, \ldots, \Delta z_{50} \}$ chosen for being representative of the ensemble of curves shown.

After $k \in \{5,10,15,\ldots,50\}$ time steps, we want to update the forecast state vector $\mathbf{x}_k^\text{f}$ with respect to the observation vector $\mathbf{y}_k$, given by
\begin{equation}
    \mathbf{y}_k = ( \Delta z_k, \ \bar{\gamma}_k )^T, \nonumber
\end{equation}
where $\Delta z_k$ is the standardised cumulative thickness after $k$ time steps, and $\bar{\gamma}_k$ is a harmonic mean of gamma ray values,
\begin{equation}
    \bar{\gamma}_k = \left( \frac{1}{n_{\gamma, k}} \sum_{z_i \in I_k} \frac{1}{\gamma(z_i)} \right)^{-1}. \nonumber
\end{equation}
The average is taken over the depth interval $I_k$, corresponding to the newest five-layer block, and $n_{\gamma, k}$ is the number of gamma ray measurements in the well log which belong to this interval. The depth interval of each block begins where the previous one ended, so that after 50 time steps, the ensemble will have been conditioned to all the well log data in the depth interval shown in the right panel of Fig. \ref{fig:Tunalik1gammaRay}.

The term $\mathbf{h}_k (\mathbf{x}_k^\text{f})$ in the update equation (\ref{pseudo_y}) entering in equation (\ref{kf_eq}) corresponds to the expected value of $\mathbf{y}_k$ given the state forecast $\mathbf{x}_k^\text{f}$ To compute
\begin{equation}
\mathbf{h}_k (\mathbf{x}_k^f) = ( \Delta z_k^\text{f}, \ \bar{\gamma}_k^\text{f} )^T \nonumber
\end{equation}
based on $\mathbf{x}_k^f$, we first extract the current thickness $\Delta z_k^\text{f}$ of the simulated layer package at the well location by taking the difference in elevation between the top and bottom surfaces. The synthetic gamma ray value for grid cell $i$ is given by
\begin{equation}
    \gamma_i^\text{f} = \sum_{\ell = 1}^4 p_{\ell, i}^\text{f} \tilde{\gamma}_\ell. \nonumber
\end{equation}
where $p_{\ell, i}^\text{f}$ is the forecasted proportion of sediment type $\ell \in \{1,2,3,4\}$ in grid cell $i$, and $\tilde{\gamma}_\ell$ is the expected gamma ray measurement for a grid cell containing only sediment of type $\ell$. If, for instance, grid cell $i'$ contains pure clay, then we will have $\gamma_{i'}^\text{f} = \tilde{\gamma}_4$. The expected gamma ray values are parameters which must be chosen in advance to calibrate the model. Here, they were chosen so that the distribution of gamma ray values obtained by simulating from the prior distribution matches the marginal, depth-averaged distribution of gamma ray measurements in the relevant part of the Tunalik 1 well log.

Once the cell-wise gamma ray values have been synthesised, we average them over the most recent five-layer block. Let the 5 top grid cells in the well, after $k$ time steps, be numbered $i_{1,k}, i_{2,k}, \ldots, i_{5,k}$. Then the synthetic gamma ray block average is
\begin{equation}
    \bar{\gamma}_k^\text{f} = \left( \frac{1}{5} \sum_{j = 1}^5 \frac{1}{\gamma_{i_{j,k}}^\text{f}} \right)^{-1}. \nonumber
\end{equation}

The covariance of the measurement noise terms, denoted $\text{Cov}(\boldsymbol\epsilon_y, \boldsymbol\epsilon_y)$ in equation (\ref{kf_eq}), must be specified. Here, we used a very large standard deviation of $\sigma_{\Delta z} = 3$ km for the cumulative thickness observations, and a standard deviation of $\sigma_{\bar{\gamma}} = 3$ API for the local gamma ray averages. The thickness and gamma ray noise terms are assumed to be independent of each other. Choosing a very large standard deviation for the thickness means we are modelling the thickness as highly uncertain. As a result, observations of thickness will not contribute much to the shape of the likelihood function, and will have a limited influence on the posterior ensemble. As the effect of compaction between the end of the simulation time period and the time of observation is not explicitly accounted for in the model, it is reasonable that the uncertainty associated with observations of depth will be much larger than the uncertainty associated with gamma ray measurements. To illustrate the effect of $\sigma_{\Delta z}$ on estimates, we also run the EnKF with $\sigma_{\Delta z} = 30$ m. 

When assessing the estimates of the system state and parameters in the North Slope case, we do not know the true state of the system, neither today nor at 115 Ma. Unlike for the simulation study in the previous section, there are no reference values of the estimated quantities to be used for judging the quality of the estimates. Still, we may get some insight by looking at the evolution of the system state estimate over time, and by comparing the synthetic data associated with the final system state estimate with the observations used for conditioning.

\begin{figure}
\begin{center}
\includegraphics[width=\textwidth]{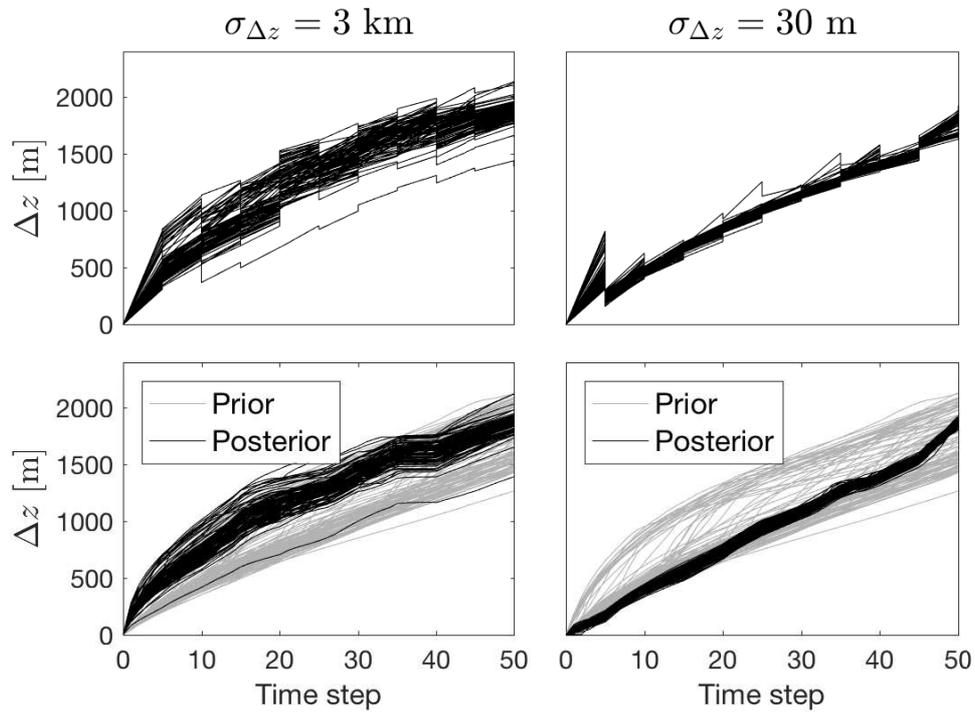}
\caption{Top: Evolution of estimate of total thickness of deposited sediment at Tunalik 1 well location. Bottom: Final analysis (posterior) ensemble compared with prior (unconditional) realisations. Results shown correspond to observing cumulative thickness with high (left) and relatively low (right) uncertainty.}
\label{fig:thicknessEvolPriorPosterior}
\end{center}
\end{figure}

The top left panel of Fig. \ref{fig:thicknessEvolPriorPosterior} shows, for the case where $\sigma_{\Delta z} = 3$ km, the evolution over the simulated time interval of total thickness at the location of the Tunalik 1 well, represented by all 100 members of every forecast and analysis ensemble. Updates occur every 5 time steps, as can be seen by the vertical shifts. The bottom left panel shows the same thickness as represented by the final analysis ensemble. Values extracted from unconditional simulations are included for reference (the same prior thickness curves are shown in the left panel of Fig. \ref{fig:Tunalik1gammaRay}). The right panels show the same for the case where $\sigma_{\Delta z} = 30$ m.

\begin{figure}
    \begin{center}
    \includegraphics[width=\textwidth]{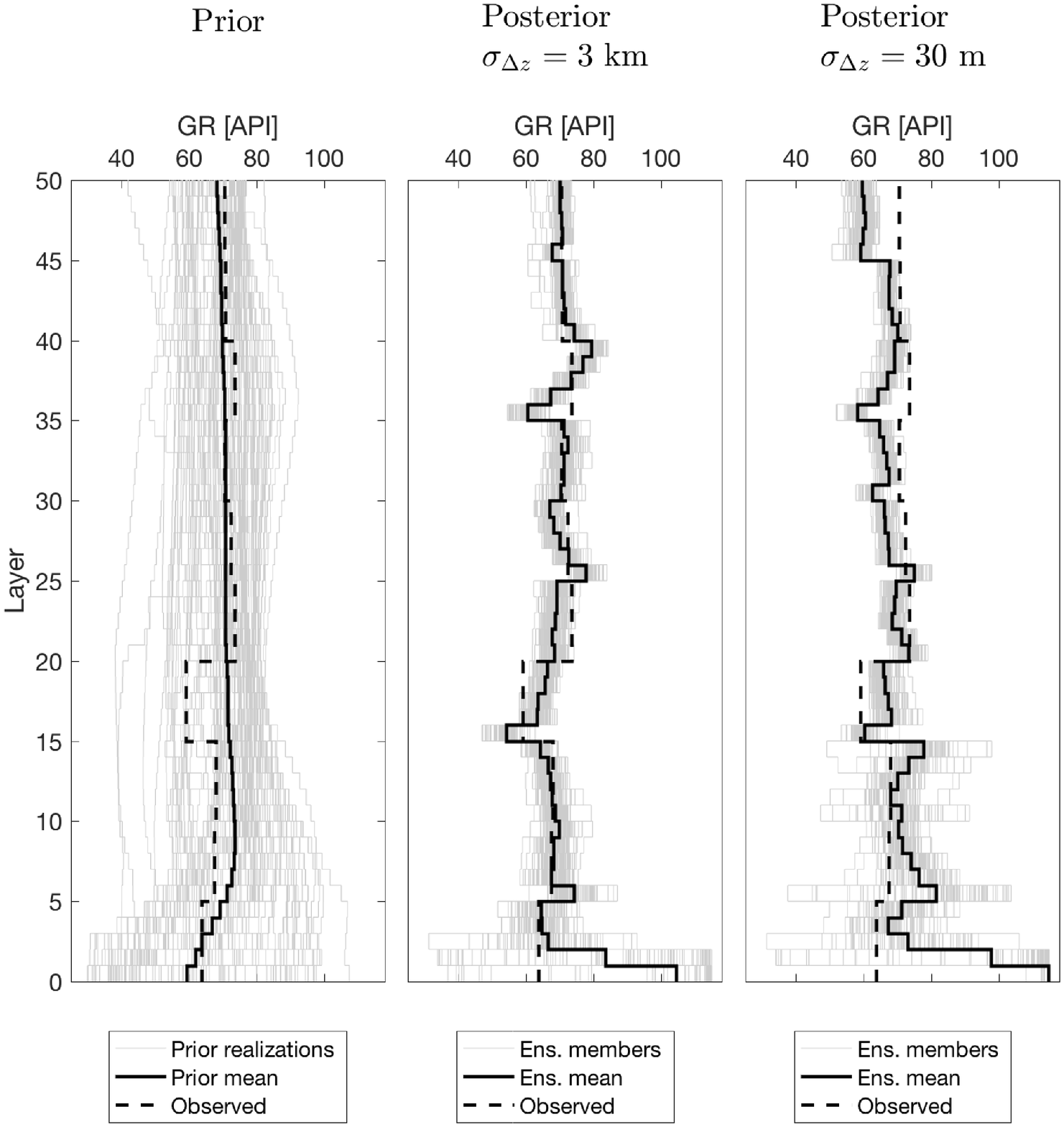}
    \caption{Synthetic and observed gamma ray measurements in the Tunalik 1 well. Left: Realisations of the prior distribution, obtained by simulating without conditioning. Middle: Ensemble members after final update at time $t_{50}$ when thickness observations are highly uncertain. Right: Posterior ensemble when thickness observations are informative. Observed gamma ray values are block-wise averages.}
    \label{fig:Tunalik1gammaRayPriorFitted}
    \end{center}
\end{figure}

Figure \ref{fig:Tunalik1gammaRayPriorFitted} shows the match between estimated and observed well log measurements. The left panel shows layer-wise gamma ray values synthesised from prior simulation results, while the middle and right panels shows gamma ray values synthesised from posterior ensemble members with large and small observation uncertainty for cumulative thickness, respectively. All three panels also include the layer-wise ensemble mean and the observed block-wise average. The latter being identical in all panels.

\begin{figure}
    \centering
    \includegraphics[width=\textwidth]{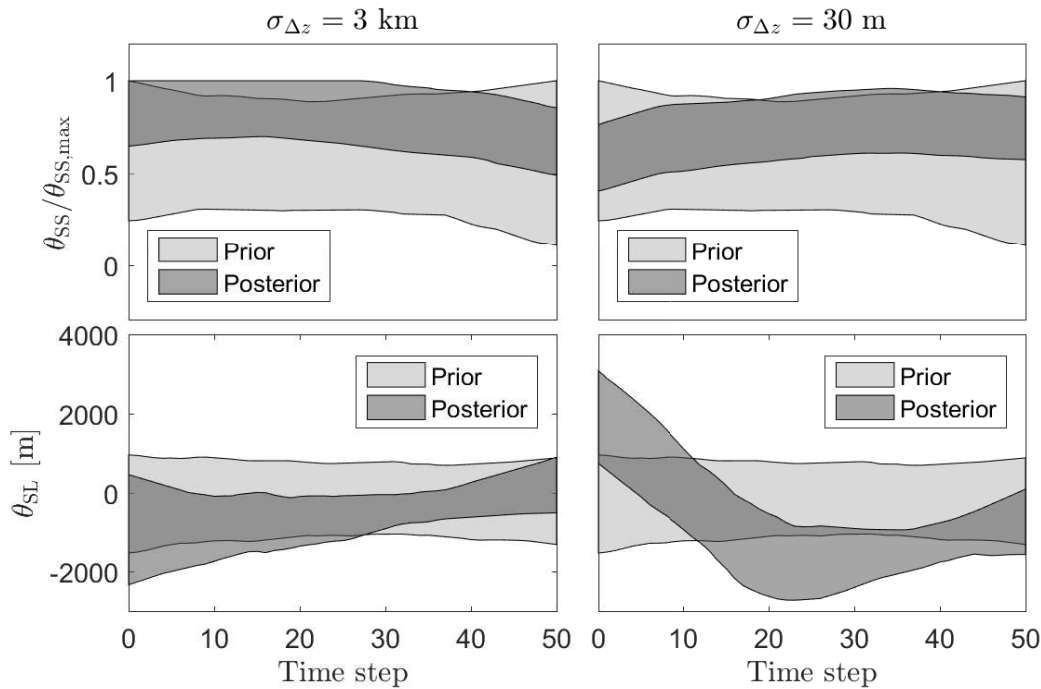}
    \caption{Prior and posterior distributions of sediment supply (top) and sea level (bottom) parameters represented by empirical point-wise 90 percent credible intervals computed from 100 realisations of each distribution. As in Fig. \ref{fig:thicknessEvolPriorPosterior}, estimates correspond to the two cases where cumulative thickness is observed with high uncertainty (left) and relatively low uncertainty (right).}
    \label{fig:NorthSlopeParameters}
\end{figure}

Figure \ref{fig:NorthSlopeParameters} shows a comparison of the prior and posterior distributions of the sediment supply (top) and sea level (bottom) parameters in the form of point-wise, empirical 90\% confidence intervals computed from ensemble members. Estimates for both large (left) and small (right) $\sigma_{\Delta z}$ are shown.

\section*{Discussion}
\subsection*{The conditioning problem}
Loosely speaking, the data assimilation task considered in this paper consists of inferring causes from partially observed results. Since the measurable outcomes of the simulated geological processes are multiply realisable in terms of the various inputs, this inference problem lacks a unique solution. In this regard, it is a typical inverse problem.

Taking a Bayesian approach is natural for two reasons. First, introducing a prior pdf for the unknown quantities to be estimated provides necessary regularisation of the solution of the inverse problem. Second, Bayesian inference is a consistent and principled way of combining quantitative observations of a physical system with relevant subject matter knowledge, while taking into account varying degrees of uncertainty associated with different sources of information \citep{Tarantola2005,Evensen2009}.
 
The EnKF is an appealing method for assimilating data to process-based numerical models like the GPM, as the sequential fashion in which the ensemble members are updated is well suited for exploiting the temporally ordered nature of the simulated sedimentation process. For situations beyond purely accumulative basin filling scenarios, however, the suitability of the EnKF is not assured. In scenarios characterised by more complex dynamical environments, for instance significant erosion events or faulting [see e.g. Chap 4.5 in \citep{Pyrcz2014}], comparing a layer forming early in the simulated time period with present-day observations may be ill-advised, since the layer could be partially eliminated or moved in a later stage of the simulation. We outline possible extensions of the data assimilation method in the discussion at the end of this paper.

\subsection*{Synthetic test case}
In the \emph{Numerical experiments} section, to provide a context for assessing the performance of the EnKF on the synthetic test case, two additional estimation methods were tested: EnS and MDA. These methods update the ensemble only at the final geological time point.

With regard to the performance measures reported in Table \ref{tab:EnKFtrialresults} and Table \ref{tab:trialresults}, it is worth keeping in mind that both MSE and CRPS depend on the scale of the variable estimated. Hence, comparing values between methods for the same variable is always valid, while comparisons between estimated variables, whether within-method or between-method, are not necessarily meaningful.

The overall conclusion to be drawn from the results of the synthetic test case is that the EnKF performs reasonably well on this problem, which has the important property that the reference realisation, or ground truth, was generated using the same simulator which was used in the estimation. Furthermore, the observations used in the conditioning were generated from the reference realisation according to the specified likelihood model. This guarantees that the prediction target is realisable by the simulator, and that the likelihood model accurately represents the data generating process. This test case, therefore, represents an ideal case, and the filter's performance here should not be expected to generalise to cases without these properties. Nevertheless, comparing the EnKF with other estimation methods on an idealised, synthetic test case, is informative of relative performance between the methods in question, at least when applied to cases with a similar structure.

Compared with the EnKF, both the EnS and the MDA perform poorly on the synthetic test case, with larger MSE and CRPS for all variables. Empirical confidence interval coverage probabilities, while a little below the mark for EnKF, are surprisingly small for both of the other estimation methods, indicating that the large linear updates that they apply to the ensemble result in underestimation of posterior uncertainty.

\subsection*{Real data test case}
In the second, more realistic test case, we model a piece of the Colville foreland basin in North Slope, Alaska, by conditioning a simulation of five million years of sedimentation on gamma ray well log data from the Tunalik 1 well, located within the basin. This is an example of the kind of basin filling scenario that we expect sequential data assimilation to be applicable to.

The depths in the Tunalik 1 well corresponding to the top and bottom of the simulated layer package were picked based on a conceptual model of the same region. Since our a priori confidence in this model is high relative to the level of uncertainty associated with the initial state and parameters, we treat the two depth markers as known constants. Nor do we attempt to explicitly model how the studied layer package changes between the end of the simulated time interval and the present day. The EnKF implementation used on the North Slope test case generally conditions on both locally averaged gamma ray measurements and observations of cumulative thickness. When the thickness observations are treated as very imprecise, by letting $\sigma_{\Delta z} = 3$ km, the system state and parameters are, in effect, being conditioned on gamma ray measurements only.

An alternative approach would be to explicitly model changes happening after the studied layer package was deposited, either by extending the simulated time period beyond the time interval of primary interest, or by using a less computationally expensive model to account for these changes. This could be a proxy model, based on a more coarse grained representation of the same processes as in the full model, or it could be a surrogate model, built by identifying regularities in the relationship between inputs and outputs of the full model [see e.g. \citep{frolov2009}]. In either case, an estimate of the present-day system state would be produced, and synthetic observations would be created by applying the likelihood model to this intermediate estimate.

The results of the real data test case are harder to interpret than the synthetic case results. Lacking a reference realisation to compare the estimates to, we resort to comparing the observations used in conditioning to predictions of the same observed quantities, synthesised from the estimated system state. In the North Slope, Alaska case, this means producing a synthetic gamma ray log from the estimated sediment proportions at the location of the Tunalik 1 well, and comparing this with the corresponding observed gamma ray measurements.

Although the data match for locally averaged gamma ray measurements in the Tunalik 1 well does leave something to be desired, it is clear, from comparing the panels of Fig. \ref{fig:Tunalik1gammaRayPriorFitted}, that both posterior ensembles fit the well log better than the prior ensemble does, with the $\sigma_{\Delta z} = 3$ km estimate achieving the closest match. For the sea level and sediment parameters, we see a marked tightening of the confidence intervals in Fig. \ref{fig:NorthSlopeParameters} going from the prior ensemble to the posterior ensembles, yet in both cases the posterior is still quite diffuse, suggesting that conditioning on gamma ray measurements from a single well yields only a modest reduction in uncertainty. The estimates of $\boldsymbol\theta_\text{SS}$ and $\boldsymbol\theta_\text{SL}$ obtained with $\sigma_{\Delta z} = 3$ km and $\sigma_{\Delta z} = 30$ m are broadly similar, the main difference being that the $30$ m estimate (bottom right panel of Fig. \ref{fig:NorthSlopeParameters}) detects a sea level decrease in the first half of the simulated time interval, which is less apparent in the $3$ km estimate (bottom left).

\subsection*{Assumptions and limitations}
When developing our problem-adapted version of the EnKF, we have assumed that the system dynamics are deterministic, so that identical inputs at one time will always produce identical outputs in the next time step. A consequence of this is that all the stochastic variation in a forecast ensemble is derived from variation in the updated ensemble one time step earlier. Adding a stochastic element to the time-evolution of the system state could be a way to account for possible model error, that is a possible discrepancy between the simulation and the actual physical processes being simulated.

As mentioned at the start of this section, we do not expect sequential data assimilation to be practical for geological scenarios deviating significantly from the kind of accumulative or additive behaviour which dominates the two test cases in this paper. Effectively conditioning simulations of more general geological processes likely requires a different approach.

\subsection*{Potential for further development}
The modifications made to the standard EnKF in this paper concern only the observation likelihood and the representation of the system state and parameters. Other modifications, affecting how covariances are estimated, and how updates are applied to ensemble members, are possible. For example, updates could be localised in time, so that layers formed recently receive a more substantial update than older layers. In many applications, localisation can have a stabilising effect on the posterior ensemble \citep{Nychka2010,Sakov2011}. Another possible modification is to inflate the variance of the ensemble for a more accurate representation of uncertainty~\citep{saetrom2013}.

With respect to extending the ensemble-based simulation conditioning approach to make it more widely applicable, two directions of extension seem especially pertinent. First, one might wish to condition a simulation to several different kinds of data at the same time. In the North Slope case, for instance, we could imagine using not just gamma ray observations, but also observations of porosity or electric potential in the same well, or we could condition the simulation to well log data from several distinct wells. We may also want to combine information from well logs with data from seismic or other geophysical surveys. What is required in either case, is a likelihood model describing how the measurable quantities relate to the unobserved system state. Given the relatively coarse lateral resolution of the North Slope case, assuming conditional independence between observations at different sites might be reasonable. If so, the task of specifying the likelihood model for a seismic survey is effectively reduced to the problem of synthesising a seismic trace at a given grid location given the system state at only that location.

The other notable direction to expand the approach in is to try and get beyond accumulative basin filling. For this to work, data cannot be assimilated sequentially, as in the straightforward implementation of the EnKF. One alternative sampling approach is Markov chain Monte Carlo sampling, see e.g. \citet{Charvin2009} or \citet{laloy2017} for applications on similar problems. It is not always clear, however how to guide such samplers to give reasonable results for complex high dimensional problems. Another possibility is to move to the particle filter (PF) or similar conditioning methods. The PF has the distinct advantage over the EnKF that it never manipulates simulator outputs directly, instead performing conditioning by updating a set of weights associated with the ensemble of model realisations. On the other hand, the PF is typically less efficient than the EnKF at sampling the space of possible parameters and states, so that a relatively large number of realisations may be needed to obtain useful estimates and to prevent weights from collapsing \citep{Chen2003}. Whether the accompanying computational cost is prohibitive or not seems a worthwhile question to pursue.

\section*{Acknowledgments}
We are grateful to Hilde G. Borgos, Bjørn Harald Fotland, Per Salomonsen, Lars Sonneland and Jan Tveiten at Schlumberger, Stavanger, Norway for access to the simulation software, technical support and helpful advice, to Oliver Schenk at Schlumberger, Aachen, Germany for much needed data and background information on the North Slope, Alaska case, and to Henning Omre at NTNU, Trondheim, Norway for useful discussions. We also thank the partners of the Uncertainty in reservoir evaluation (URE) project at NTNU.

\section*{Code Availability}
Matlab scripts used to carry out computations, analyse results, and create some of the figures in this article are available online at github.com/Skauvold/DA-GPM (DOI: 10.5281/zenodo.1012346).

\section*{Conflict of Interest}
No conflict of interest declared.

\bibliography{gpmenkf}

\clearpage
\section*{Appendix: Implementation details of the EnKF}
In our situation, the EnKF is used to build a sequential approximation to the conditional probability density function (pdf) of the geological state variables, given information from the well log.

\begin{figure}
\begin{center}
\includegraphics[scale=1]{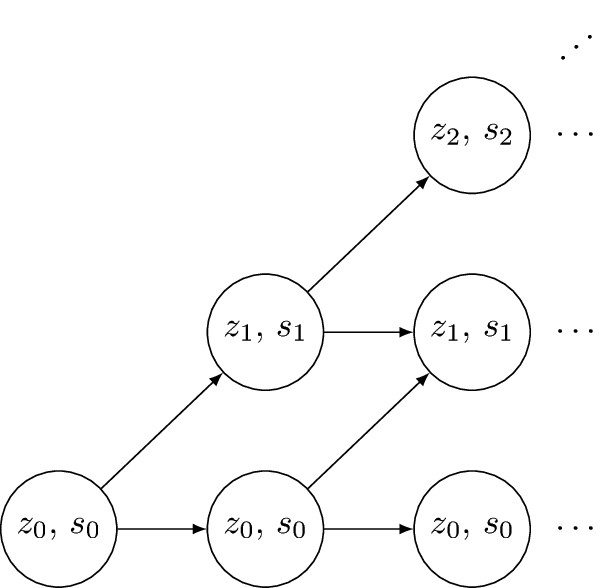}
\caption{Graph illustrating how the dimension of the state variable increases with each time step.}
\label{fig:graphEXT}
\end{center}
\end{figure}

Let $\mathbf{v}_{k}$ be the state vector of variables at geological time $t_k$, $k \in \{1,\ldots,n_t\}$. This state is constructed from two distinct parts: 
a) the layer elevations $\mathbf{z}_k$ and the layer sediment compositions $\mathbf{p}_k$, b) the sediment supply $\boldsymbol\theta_\text{SS}$ and sea level $\boldsymbol\theta_\text{SL}$. Parts in a) are layer variables represented on a grid of lateral coordinates, while part b) variables do not vary with location. The elevations and sediment compositions in part a) are connected over geological time by the GPM forecast model, so that with $k' < k$, we have
\begin{equation}\label{dynam}
(\mathbf{z}_k,\mathbf{p}_k) =  F ( \mathbf{z}_{k'},\mathbf{p}_{k'},\boldsymbol\theta_\text{SS},\boldsymbol\theta_\text{SL}).
\end{equation}
The sediment proportions $\mathbf{p}_k$ have a one-to-one relation with another variable $\mathbf{s}_k$, in the form of a logistic transformation \citep{dobson2008}. The transformation is identical for all grid cells. Considering one time step and one grid cell only, and the four sediment types, we have 
$s_j = \log \left( \frac{p_j}{p_4} \right)$, $j=1,2,3$, with inverse transformation $p_j=\frac{e^{s_j}}{1+\sum_{j=1}^3 e^{s_j}}$ and $p_4=\frac{1}{1+\sum_{j=1}^3 e^{s_j}}$. Here, $p_j \geq 0$, $j=1,2,3,4$, and $\sum_{j=1}^4 p_j = 1$, while $s_j \in (-\infty,\infty)$, $j=1,2,3$, which makes it more robust to the linear updating in the EnKF.
Layers are built up over geological time according to equation (\ref{dynam}), and we include in the state vector all layers generated up to the current time (Fig. \ref{fig:graphEXT}). Part b) variables are represented as curves indexed by time, and the entire curve (for the whole simulated geological time interval) is included in every state vector. There is hence no change in part b) variables in the forecast step.
Altogether, the state vector at time $k$ is then 
$$
\mathbf{v}_{k}=(\mathbf{z}_0,\mathbf{s}_0,\ldots,\mathbf{z}_k,\mathbf{s}_k,\boldsymbol\theta_\text{SS},\boldsymbol\theta_\text{SL}).
$$

The EnKF is based on Monte Carlo sampling. At the initial time, $n_e$ ensemble members are sampled independently from the prior pdf $p(\mathbf{z}_0,\mathbf{s}_0,\boldsymbol\theta_\text{SS},\boldsymbol\theta_\text{SL})$. For later time steps $k$, the EnKF consists of two steps: i) the forecast step and ii) the analysis step (also referred to as the update step).
In i) the state vector is propagated forward in geological time as described above. We denote the $n_e$ members of the forecast ensemble by $\mathbf{v}^{1,\text{f}}_k$, $\ldots$, $\mathbf{v}^{n_e,\text{f}}_k$. The assimilation in ii) is done by building a regression model between the state variables and the data, and then using a linear update formula:
\begin{equation}
\mathbf{v}^{b,\text{a}}_k = \mathbf{v}^{b,\text{f}}_k + \hat{\mathbf{K}}_k(\mathbf{y}_k-\mathbf{y}^{b}_k), \quad \hat{\mathbf{K}}_k = \hat{\boldsymbol\Sigma}_{vy,k} \hat{\boldsymbol\Sigma}_{y,k}^{-1}. \nonumber 
\end{equation}
Here, the $\mathbf{y}^{b}_k$ are pseudo-data obtained from $\mathbf{v}^{b,\text{f}}$ and the likelihood model, while $\hat{\boldsymbol\Sigma}_{vy,k}$ and $\hat{\boldsymbol\Sigma}_{y,k}$ are the empirical cross-covariance and covariance matrices of the data:
\begin{align*}
\hat{\boldsymbol\Sigma}_{y,k} &= \frac{1}{n_e} \sum_{b=1}^{n_e} (\mathbf{y}^{b}_k-\bar{\mathbf{y}}_k) (\mathbf{y}^{b}_k-\bar{\mathbf{y}}_k)^T, \quad &\bar{\mathbf{y}}_t=\frac{1}{n_e} \sum_{b=1}^{n_e} \mathbf{y}^{b},\\
\hat{\boldsymbol\Sigma}_{vy,k} &= \frac{1}{n_e} \sum_{b=1}^{n_e} (\mathbf{v}^{b,\text{f}}_k-\bar{\mathbf{v}}^\text{f}_k) (\mathbf{y}^{b}_k-\bar{\mathbf{y}}_k)^T, \quad &\bar{\mathbf{v}}^\text{f}_k=\frac{1}{n_e} \sum_{b=1}^{n_e} \mathbf{v}^{b,\text{f}}.
\end{align*}

In our context, the likelihood model is $\mathbf{y}_k = \mathbf{h}_k (\mathbf{v}_k)+\boldsymbol\epsilon_{y,k}$, where $\boldsymbol\epsilon_{y,k}$ is a zero-mean Gaussian vector with covariance matrix $\text{Cov}(\boldsymbol\epsilon_{y},\boldsymbol\epsilon_{y})$, which gives the form described in the main body of this paper [equation (\ref{kf_eq})], with
\begin{align*}
&\widehat{\mbox{Cov}}[\mathbf{v}_k^\text{f},\mathbf{h}_k (\mathbf{v}_k^\text{f})]=\frac{1}{n_e} \sum_{b=1}^{n_e} [\mathbf{v}^{b,\text{f}}_k-\bar{\mathbf{v}}^\text{f}_k] [\mathbf{h}_k (\mathbf{v}_k^{b,\text{f}})-\bar{\mathbf{h}}_k (\mathbf{v}_k^\text{f})]^T, \quad \bar{\mathbf{h}}_k (\mathbf{v}_k^\text{f})=\frac{1}{n_e} \sum_{b=1}^{n_e} \mathbf{h}_k(\mathbf{v}^{b,\text{f}}),\\
&\widehat{\mbox{Cov}}[\mathbf{h}_k (\mathbf{v}_k^\text{f}),\mathbf{h}_k (\mathbf{v}_k^\text{f})]=\frac{1}{n_e} \sum_{b=1}^{n_e} [\mathbf{h}_k (\mathbf{v}_k^{b,\text{f}})-\bar{\mathbf{h}}_k (\mathbf{v}_k^\text{f})]  [\mathbf{h}_k (\mathbf{v}_k^{b,\text{f}})-\bar{\mathbf{h}}_k (\mathbf{v}_k^\text{f})]^T.
\end{align*}

After the final analysis step the ensemble represents an approximation of the posterior pdf of the geological variables, given all data.
For Gauss-linear dynamical systems, the EnKF is asymptotically correct, and the approximation becomes exact in the limit as $n_e \rightarrow \infty$. For other situations, there are no theoretical results regarding the quality of the approximation, but it has shown very useful in many practical applications.
\end{document}